\documentclass{pnastwo}

\usepackage{graphicx}
\usepackage{amssymb,amsfonts,amsmath}

\url{www.pnas.org/cgi/doi/10.1073/pnas.0709640104}
\copyrightyear{2008}
\issuedate{Issue Date}
\volume{Volume}
\issuenumber{Issue Number}
\begin{document}
\title{Point process analysis of large-scale brain fMRI dynamics.}

\author{Enzo Tagliazucchi\affil{1}{Departamento de F\'isica, Facultad de Ciencias Exactas y Naturales, Universidad de Buenos Aires, Argentina.},
Pablo Balenzuela\affil{1}{}\affil{2}{Consejo Nacional de Investigaciones Cient\'ificas y Tecnol\'ogicas (CONICET), Buenos Aires, Argentina},
Daniel Fraiman\affil{2}{}\affil{3}{Departamento de Matem\' atica y Ciencias, Universidad de San Andr\' es, Buenos Aires, Argentina.} 
\and
Dante R. Chialvo\affil{2}{}\affil{4}{David Geffen School of Medicine, UCLA, Los Angeles, CA. USA.} \affil{5}{Facultad de Ciencias M\'edicas, Universidad Nacional de Rosario, Rosario, Argentina.}
}

\contributor{Submitted to Proceedings of the National Academy of Sciences
of the United States of America}

\maketitle

\begin{article}
\begin{abstract}
Functional magnetic resonance imaging (fMRI) techniques have contributed significantly to our understanding of brain function. Current methods are based on the analysis of \emph{gradual and continuous} changes in the  brain blood oxygenated level dependent (BOLD) signal. Departing from that approach, recent work has shown that equivalent results can be obtained by inspecting only the relatively large amplitude BOLD signal peaks, suggesting that relevant information can be condensed in \emph{discrete} events. This idea is further explored here to  demonstrate how brain dynamics at resting state can be captured just by the timing and location of such events, i.e., in terms of a spatiotemporal point process.  As a proof of principle, we show that the resting state networks (RSN) maps can be extracted from such point processes. Furthermore, the analysis uncovers avalanches of activity which are ruled by the same dynamical and statistical properties described previously for neuronal events at smaller scales. Given the demonstrated functional relevance of the resting state brain dynamics, its representation as a discrete process might facilitate large scale analysis of brain function both in health and disease.

\end{abstract}

\keywords{fMRI | criticality | brain dynamics | point processes}

%\abbreviations{SAM, self-assembled monolayer; OTS, octadecyltrichlorosilane}

\dropcap{I}mportant efforts to understand brain function, both in health and disease, are concentrated in the analysis of large-scale spatiotemporal patterns of brain activity available from fMRI techniques \cite{raichle,greicius,fox2007,smith,beckmann2004,beckmann2005}, allowing for instance the unravelling of the functional connectivity between all possible brain regions, as is done under the Connectome project \cite{conectome,conectome2,conectome3}. Novel techniques of analysis are needed because the difficulty of managing extremely large data sets is increased by new advances in imaging technology continuously improving temporal and spatial resolution. In that direction recent work has shown that important features of brain functional connectivity at rest can be computed from the relatively large amplitude BOLD fluctuations \cite{enzo2} after the signal crosses some amplitude threshold.
Pursuing the same general idea, it would be desirable to explore complementary ways of data reduction. In particular we are interested in a  method often used to study the structure and properties of attractors of dynamical systems, which consists in the introduction of a Poincar\`e section. By definition, this approach decreases the dimension of the phase space, and consequently the size of the data sets, facilitating in this way further numerical investigations. In general, there exist two possibilities: the first one is to analyze the set of points which are the coordinates of  the successive intersections of the secant Poincar\`e plane by the phase space trajectories. The second possibility is to study the series of time intervals between the consecutive intersections. The resulting time intervals constitute a so-called point process \cite{cox}, a construction useful in many areas of science, including neuroscience. It has been shown  that  under certain conditions  the most important statistical features of the dynamical regime can be condensed into a point process \cite{poincare1,poincare2,poincare3,poincare4,poincare5,poincare6}.

The motivation to attempt similar approach in fMRI data is strengthened by the observation that, in response to neuronal activation, the BOLD signal often repeats a stereotypical pattern \cite{enzo2, friston1995,friston1998,aguirre}. This feature of the BOLD response dynamics suggests that it should be possible to compress the data sets using the temporal marks of some Poincare section of the BOLD signal.  This is the hypothesis explored here, which implies that, in principle, the entire brain resting state functional connectivity can be reconstructed solely on the basis of the time and location of the BOLD signal threshold crossings. Besides its practical importance for fMRI signal processing, this approach may provide further clues on the dynamic organization of the brain RSN.

The paper is organized as follows: first the definition of the point process is sketched, followed by its validation by replicating well studied aspects of the fMRI brain resting state. The spatiotemporal statistics are then considered, revealing novel aspects of the brain dynamics which are scale invariant, consistent with that shown for other systems at the critical state \cite{bak,jensen,chialvo2010,expert}.

 %%%%%%%%%%%%%%%%%%%%%%%%%
 \begin{figure}[h]
\centerline{
 \includegraphics[width=.5\textwidth]{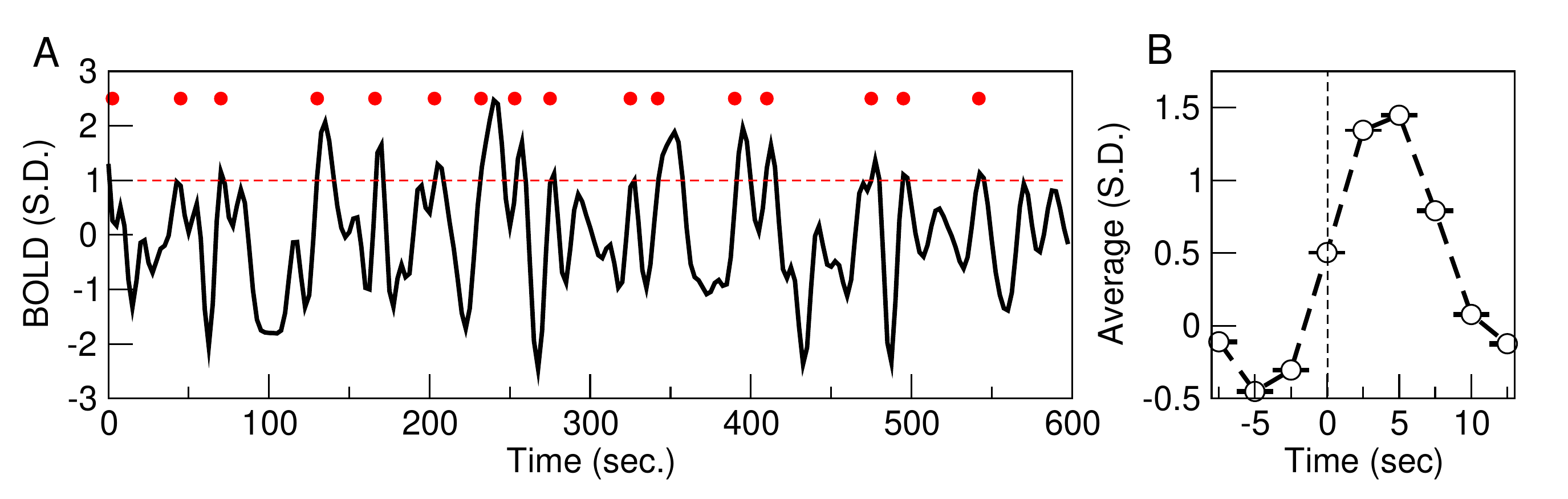}}
 \caption{(A) Example of the normalized BOLD signal of one voxel and its point process (filled circles) defined by the threshold (dashed line at 1 S.D.) crossings events. (B) Average BOLD signal (for one subject) triggered at each threshold crossing as in \cite{enzo2}. } 
 \end{figure}
\section{Results}
 
The fMRI dataset is reduced to a spatiotemporal  point process by first normalizing each BOLD signal by its own S.D., and subsequently selecting the time points at which the increasing signal crosses a given threshold (1 S.D. in this case) as it is shown in the example of Figure 1.  Notice that  the average BOLD signal around the extracted points (Figure 1B) resembles the typical hemodynamic function seen in response to an stimulus \cite{friston1995,friston1998}, despite the fact that in this case there are not explicit inputs, since recordings were obtained under resting conditions.

The timing and spatial location of the extracted points is the only information needed for further analysis. For the parameter used here, from each voxel  BOLD time series (240 samples) only $15 \pm 3$ points are threshold crossings (about one point every 40 sec) which corresponds to near 94\% reduction of the data. (see additional estimations in Supp. Info).
\begin{figure}[h]
\centerline{
 \includegraphics[width=.5\textwidth]{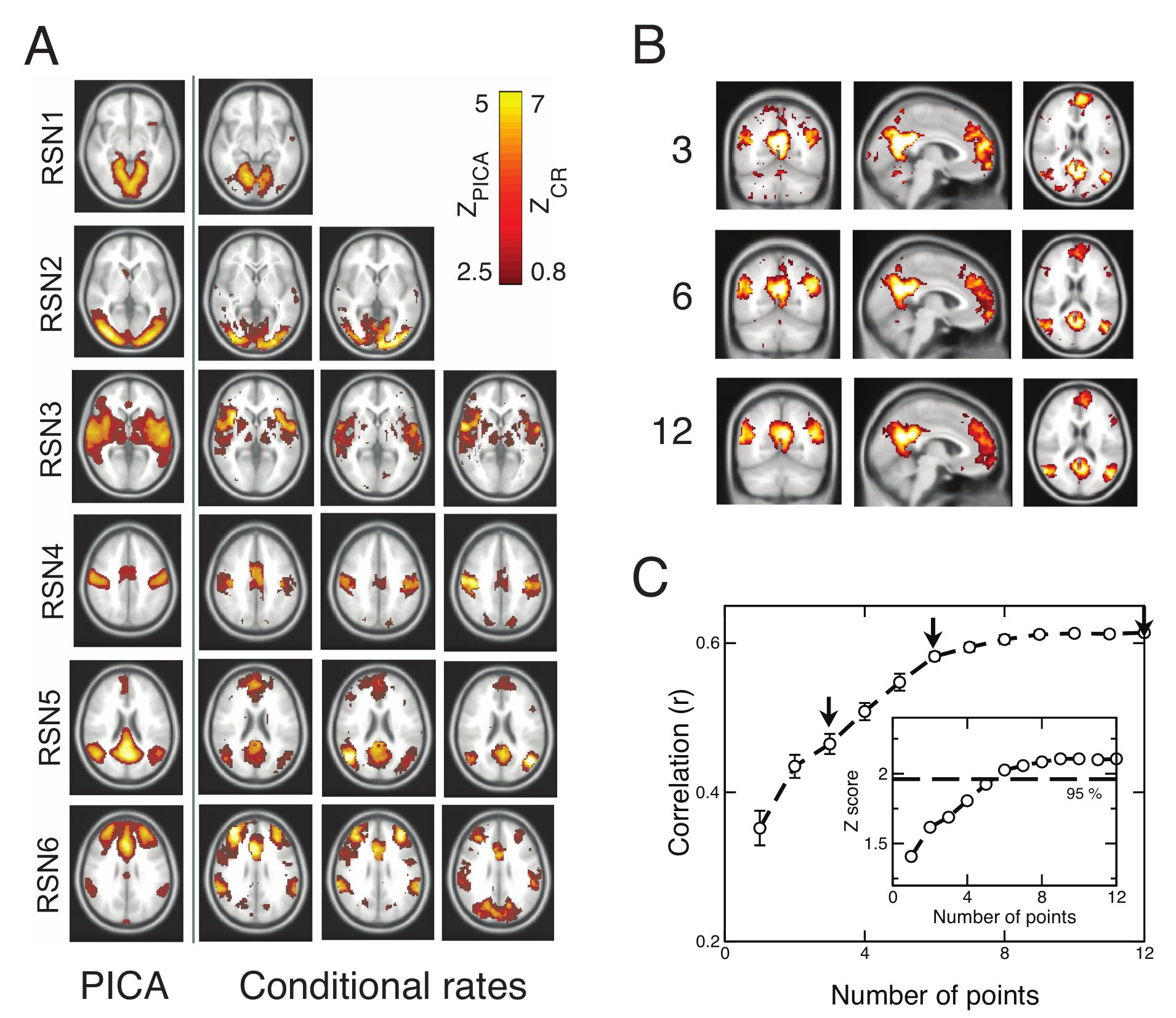}}
 \caption{RSN maps constructed with the point process compared with standard PICA. (A) PICA spatial maps (left column) and rate of points conditional to activity at a given seed (rightmost three columns, each one corresponds to a different seed).  (Slice z coordinates are -12, 0, 0, 36, 20, 26 for RSN 1 to 6; for coordinates see Table SI).  Scales for PICA ($Z_{PICA}$) and conditional rate ($Z_{CR} $) calculations are depicted in the inset.
(B) Conditional rate maps constructed using 3, 6  and 12 events of the point process at the ANGL seed (averaged from ten subjects.
Slice coordinates are x=-4,y=-60,z=18)
(C) Correlation between RSN5 (the default-mode network, DMN) PICA-derived map and the point process-derived conditional rate maps, as a function of the number of points used.  Arrows denote the examples of panel B. Z scores (number of points as degrees of freedom) with the line of 95\% confidence are plotted in the inset.  } 
 \end{figure}

{\bf Resting state networks maps derived from the point process.}
 Despite the very large data reduction, we found that the information content of  the few remaining points is very high. To demonstrate this we use the point process to calculate the location of six well known RSN maps. The computation of these maps are compared with those obtained from the full BOLD signal using a well established method (probabilistic independent component analysis - PICA \cite{beckmann2005}). 
This is done by calculating in six RSNs the rate of points co-occurrence  (up to 2 time units later in this case) between representative sites (``seeds'') and all other brain voxels and presented as maps in Figure 2. The seeds locations were selected according with previous work (see coordinates in Table SI). The similarities between our conditional rate maps and the respective PICA maps (rightmost three columns and left column of Figure 2A respectively) is already obvious  to the naked eye and confirmed by the correlation plotted in Fig. 2C. The calculation shows that despite using less than $6\%$ of the raw fMRI information,  about 5 points (on average) are enough to obtain RSN maps that are  highly correlated (95\% confidence) with those obtained using PICA of the full BOLD signals.

{\bf Structure and dynamics of the active clusters.}
The analysis of the brain large scale spatiotemporal patterns using the full BOLD signals is hampered by numerical constraints. In contrast,  the analysis of brain dynamics with the derived point process is simpler by compiling the statistics of clusters of points, both in space and time. Clusters are groups of contiguous voxels with signal above the threshold at a given time, identified by a scanning algorithm in each fMRI volume.  Figure 3A shows examples of clusters (in this case non-consecutive in time) depicted with different colors. 
Typically (Fig. 3B top) the number of clusters at any given time varies only an order of magnitude around the mean ($\sim 50$). In contrast, the size of the largest active cluster fluctuates widely, spanning more than four orders of magnitude.

The analysis reveals four novel dynamical aspects of the cluster variability which hardly could has been uncovered with previous methods. 1) At any given time, the number of clusters and the total activity (i.e., the number of active voxels) follows a non-linear relation resembling that of percolation \cite{bak}. At a critical level of global activity ($\sim 2500$ voxels, dashed horizontal line in Fig. 3B, vertical in Fig. 3C) the number of clusters reaches a maximum ($\sim 100-150$) as well as its variability. 2) The correlation between the number of active sites (and index of total activity)  and the number of cluster reverses above a critical level of activity, a feature already described in other complex systems anytime that some increasing density competes with limited capacity \cite{bak}.
3) The rate at which the very large clusters (i.e., those above the dashed line in 3B) occurs ($\sim$ one every 30-50 sec) corresponds to the low frequency range at which RSN are typically detected using PICA \cite{beckmann2004,beckmann2005}. 4) The distribution of sizes (Figure 3D) reveals a scale free distribution.  
%%%%
\begin{figure}
\centerline{
 \includegraphics[width=.5\textwidth]{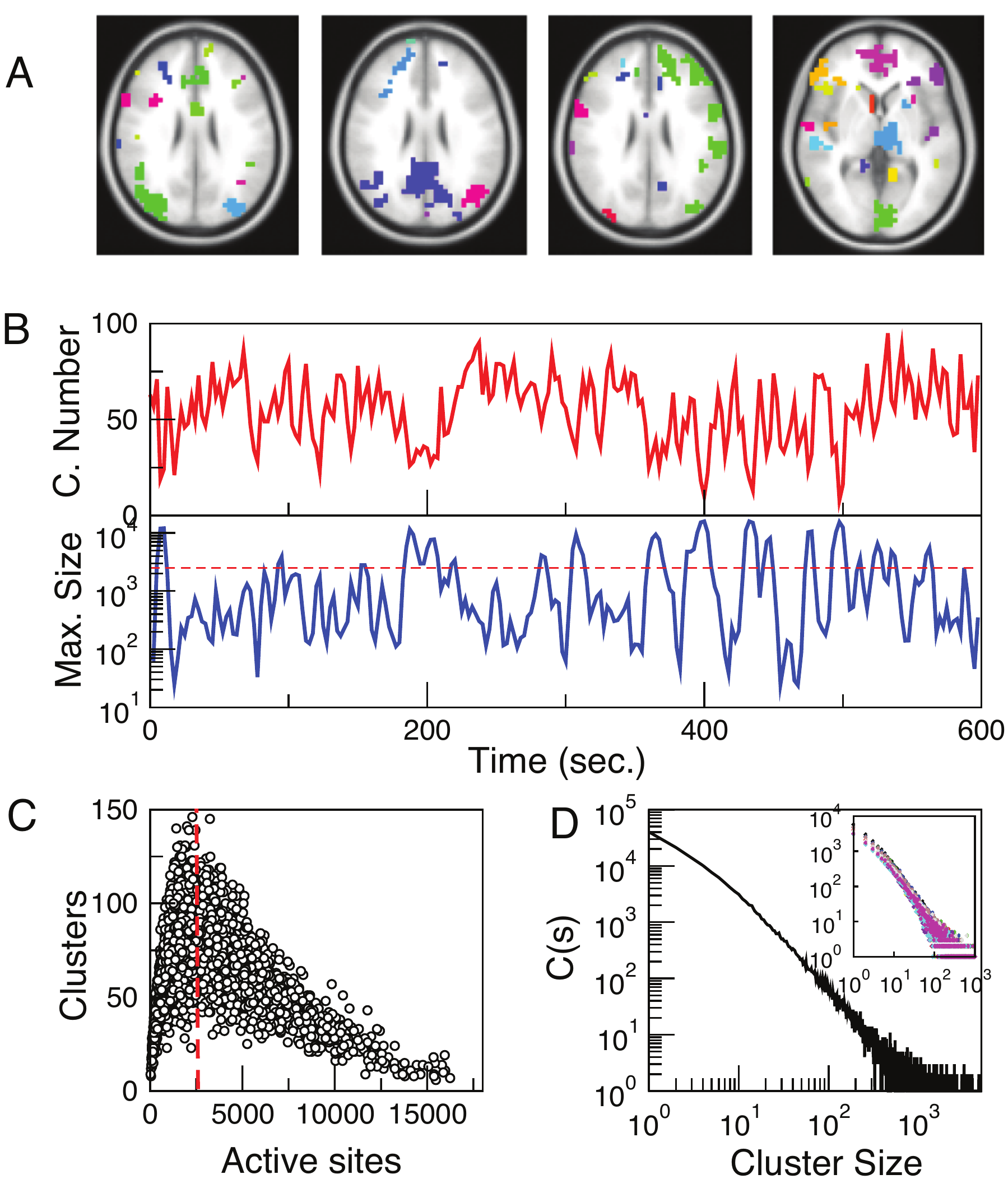}}
 \caption{Typical fluctuations of the clusters. (A) Examples  of co-activated clusters of neighbors voxels (depicted in different colors). (B) Example of the temporal evolution of the clusters' number and maximum size (in units of voxels) in one individual. (C) Instantaneous relation between the number of clusters vs. the number of active sites showing a positive/negative correlations depending whether activity is below/above a critical value ($\sim 2500$ voxels, indicated by the dashed line here and in Panel B). (D) The cluster size distribution, follows an inverse power law spanning four orders of magnitude. Individual statistics for each of the ten subjects are plotted in the inset. } 
 \end{figure}
 %%%%%%
These four features are known to be generic properties of other complex systems \cite{bak,jensen,chialvo2010} and as such it should be further explored wether they are preserved under developmental changes, aging and pathological conditions.   Notice, for instance, that the nonlinear correlation (i.e. Fig. 3 B and C) between the number of clusters and the size of the largest activated cluster can be related (and an objective measure) to the balance between excitation and  inhibition, a property believed to be fundamental for healthy cortical function. In the same direction, the size of the largest cluster and the number of clusters can be seen in terms of the amount of integration and segregation long advocated by Tononi et al. \cite{tononi,sporns} as the fundamental conundrum that the healthy cortex needs to be executing at any given time. The important issue to be noted here is that the point process approach allows for the first time a straightforward  quantification of these properties. 
%%%%%%%%%%
\begin{figure}[h]
\centerline{
 \includegraphics[width=.5\textwidth]{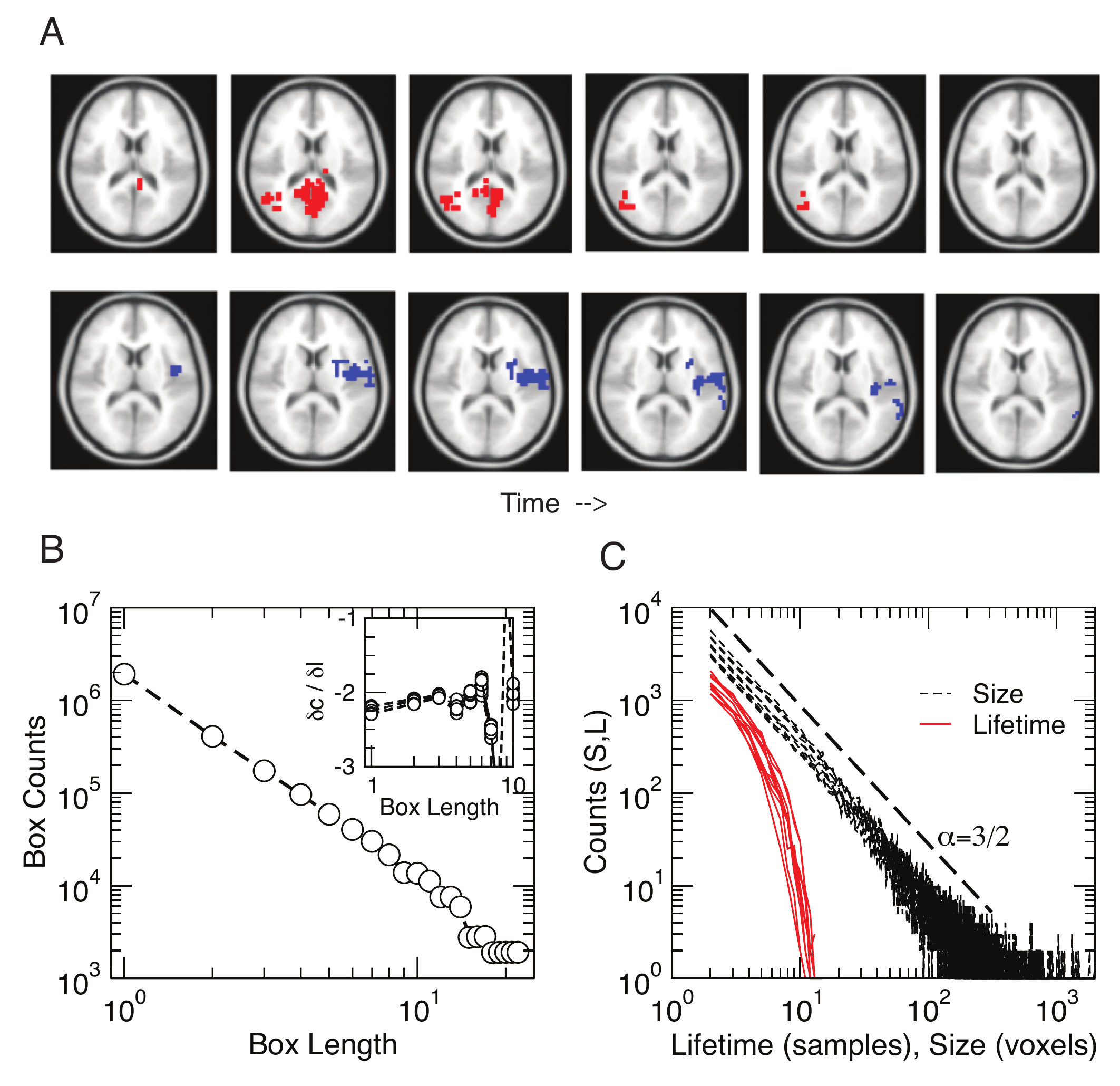}}
 \caption{ Clusters spread throughout the brain as scale-free avalanches. (A) Two examples of avalanches, one triggered from the visual cortex (top) and another from insular cortex (bottom). (B) Average cluster's fractal dimension $D\sim2.15\pm0.02$ estimated by the slope of the counts vs. length plot. Inset: derivatives between points in the main plot for each subject.
 (C) Avalanche's size and lifetime distribution function computed from about 8000 avalanches in each of 10 subjects. While avalanche size follows an inverse power law, their lifetimes density decrease faster.  The dashed line corresponds to an exponent of 3/2, found previously for neuronal avalanches \cite{plenz}. } 
 \end{figure}
%%%%%%%%%%%%
{\bf Activity spread is scale-free.}
Now we turn to explore in detail how clusters evolve in space. This is motivated by the usual notion that cognition and behavior underlying mechanism involves the sequential activation of clusters of coherent events through regions of the cortex. In that sense, we track over time an activated cluster which can appear, grow to achieve a maximum size and then disappear (or translate or divide into sub-clusters) as shown in the example of Fig. 4A.  the data reduction allow us to characterize easily this process, and in particular study two properties. For each cluster, first we measured a static space filling property,  the average fractal dimension $D$. This is shown in Fig. 4B which illustrate that $D \sim 2.15\pm 0.02 $. Second we look at the dynamics of the cluster propagation, which we found happens in bursts. The statistics in Fig. 4C  shows that avalanche's could last up to 30 sec. with sizes up to $10^3$ and have no preferred scale, being fitted to an inverse power law with the same $3/2$ exponent described previously in smaller scales \cite{plenz,peterman,chialvo2010}.
When we look at the spatial coverage of typical avalanches at their maximum size, we found that is not arbitrary, because they remain confined to the RSN' contiguous regions at which they started. For instance, in the RSN5 (i.e., DMN) an avalanche could start at the Medial Prefrontal Cortex (mPFC) would cover the entire mPFC, but will not extend to the Angular Gyrus which is consistent with the notion of RSN being independent components (see Supp. Info.). 

\section{Discussion}
As far as we know, this is the first attempt to describe large scale brain dynamics as a point process. The only previous report we are aware of \cite{jensen2007} dealt with the reverse process: how to model the continuous fMRI signal starting from a spatiotemporal point process.
 
Concerning the reason behind the surprisingly successful data reduction achieved, we can only conjecture that the upward going BOLD signals are nonlinear events where the times of crossings preserve the most relevant information. This could be in line with recent findings of all-or-none ``coherence potentials'', exhibiting avalanches with identical scaling behavior \cite{plenz3} in monkey cortex. In a similar fashion, the reduction of the fMRI BOLD signal to discrete events not only allows the identification of well-described resting state networks but also it is shown to organize in neuronal avalanches. 

The fact that avalanches with the same  power law are observed at a large scale is consistent with the hypothesis that brain dynamics operates at a critical point of a second order phase transition \cite{chialvo2010}, thus rendering the system scale invariant and allowing its description at all scales by the same set of statistical relations.  The spatio-temporal scale invariance shown in  the findings of this work and previous electrophysiological experiments on neuronal avalanches is further supported by the observation that  the correlation function of fMRI BOLD signals exhibits fractal properties \cite{expert} and the correlation length of activity measured with fMRI diverges as predicted by the theory of phase transitions \cite{fraiman2010}.

The observation that large scale brain dynamics can be traced as discrete events (scale free avalanches of activity) raises the question of the physiologically relevance encoded in the timing of these events. For instance, although rare, avalanches in the tail of the power law distribution emerge from a local origin and propagate as far as the length of the entire cortex, suggesting 
a role in the binding processes of far apart cortical regions. Additionally, the nonlinear relation  between activated cortical tissue and number of clusters exhibits an optimal point, in which the highest levels of brain activity are segregated into the maximum number of spatially isolated activations. It is interesting to investigate whether total or partial disruption of these large events, as well as alterations in the balance between activation and segregation into clusters could be correlated with pathological conditions and  with the level of awareness of the subject.

The present results are  consistent with observations at faster time scales, in which  the scalp EEG is reduced to a certain number of stereotypical topographical maps (i.e., EEG microstates) \cite{koenig} and the observation of non-stationarities with define discrete segments of electrical activity \cite{kaplan}.  Both descriptions of electrical scalp activity have also been shown to exhibit properties consistent with critical dynamics \cite{vandeville, allegrini}.  Further multimodal imaging studies could link these observations together in the context of discrete  avalanches of neural activity propagating through the cortex and determine their functional relevance for health and disease.

In summary, the results show  that the location and timing of the largest BOLD fluctuations define a spatial point process containing   substantial information of the underlying brain dynamics. Despite the very large data reduction ($> 94\%$), the approach was validated by the favorable comparison of the conditional rate maps of  avalanching activity  with those constructed with the full fMRI BOLD signals using PICA. In addition to uncover new dynamical properties for the activated clusters the method exposed scale-invariant features conjectured in the past \cite{chialvo2010} which are identical to those seen at smaller scales \cite{plenz,peterman,chialvo2010}. Beyond its potential value for fMRI signal processing, the ability of the present approach to capture relevant brain dynamics underscoring nonlinear aspects of the BOLD signal deserve further exploration.

\section{Material and Methods}
{\bf fMRI data acquisition and preprocessing.} Data was obtained, after informed consent, from ten right-handed healthy volunteers (9 female, 1 male; mean age=49, S.D.=12) scanned each 2.5 sec. during 10 minutes, requested to keep their eyes closed and to avoid falling asleep (see Supp. Info for scanning parameters). The study was approved by the Clinical Research Ethics Committee of the University of the Balearic Islands, (Palma de Mallorca, Spain). The Melodic package was used for the calculation of ICA of the RSN \cite{beckmann2004} in (Fig. 2) as well as for de-noising motion artifacts. 

{\bf Cluster and avalanche analysis.}
Spatial clusters of activated voxels were identified using an algorithm implemented in MATLAB, based on the detection of connected components in a co-activated first neighbors graph (see Supp. Info for a description of the algorithm). Clusters' fractal dimension was calculated using a standard box-counting algorithm. Avalanches were defined (similarly to that done in sandpile models, and others \cite{bak,jensen})  as starting with the isolated activation (i.e., not by any of its neighbors) of a previously inactive voxel (or group of voxels), continuing while at least one contiguous voxel is active in the next time step and otherwise ends. The avalanche tracking algorithm implemented in this work uses as a criteria for avalanche membership non empty intersection with a previously identified cluster of the avalanche at a previous time. This is able to resolve shrinking and expanding of clusters, translation and division, whenever there is spatial overlap at subsequent times. For a technical description of the algorithm, see Supp. Info.

 %Put methods in here.  If you are going to subsection it, use \verb|\subsection| commands.  Methods section should be less than 800 words and if it is less than 200 words, it can be incorporated into the main text.

%% Put the bibliography here, most people will use BiBTeX in
%% which case the environment below should be replaced with
%% the \bibliography{} command.

\begin{thebibliography}{999}

\bibitem{raichle}Raichle, M.E. The brainÕs dark energy. \emph {Science} {\bf 314},1249--1250 (2006).
 
 \bibitem{greicius}Greicius MD, Krasnow B., Reiss, A.L., Menon, V. Functional connectivity in the resting brain: a network analysis of the default mode hypothesis. \emph{Proc Natl Acad Sci USA} {\bf 100}, 253--258 (2003).

\bibitem{fox2007}Fox, M.D. \& Raichle, M.E. Spontaneous fluctuations in brain activity observed with functional magnetic resonance imaging. \emph{Nat Rev Neurosc} {\bf 8}, 700--711 (2007).

\bibitem{smith}Smith, S.M., Fox, P.T., Miller, K.L., Glahn, D.C., Fox, P.M., Mackay, C.E., Filippini, N., Watkins, K.E., Toro, R., Laird, A.R., Beckmann, C.F. Correspondence of the brain's functional architecture during activation and rest.  \emph{Proc Natl Acad Sci USA} {\bf 106}, 13040--13405 (2009).

\bibitem{beckmann2004}Beckmann, C.F. \& Smith, S.M. Probabilistic independent component analysis for functional magnetic resonance imaging. \emph{IEEE Trans Med Imag} {\bf 23},137--152 (2004).

\bibitem{beckmann2005}Beckmann, C.F., DeLuca, M., Devlin, J.T., Smith, S.M. Investigations into resting-state connectivity using independent component analysis.  \emph{Phil Trans R Soc B} {\bf 360} ,1001--1013 (2005).

\bibitem{conectome} Sporns O., Tononi, G., Kotter R. The human connectome: a structural description of the human brain. \emph{PLoS Computational Biology} {\bf 1}, 245--251 (2005).

\bibitem{conectome2} Sporns O. The human connectome: a complex network, \emph{ Annals of the New York Academy of Sciences}  {\bf 1224}, 101--125 (2011).

\bibitem{conectome3} http://www.humanconnectomeproject.org/





\bibitem{enzo2}Tagliazucchi, E., Balenzuela, P., Fraiman, D., Montoya, P., Chialvo, D.R. Spontaneous BOLD event triggered averages for estimating functional connectivity at resting state, \emph{Neurosc. Lett.}. {\bf 488}, 158--163. (2010).

\bibitem{cox} Cox D.R., \& Isham, V. (1980). Point Processes. (London and New York: Chapman and Hill).
\bibitem{poincare1}Grassberger P. \& Procaccia, I.\emph{Physica} {\bf9D}, 189 (1983).

\bibitem{poincare2} Packard, N.,  Crutchfield, J.,  Farmer, J.D. , Shaw, R. \emph{Phys. Rev. Lett.} {\bf 45}, 712 (1980).
\bibitem{poincare3}Roux, J.C., Rossi, J., Bachelart, S.,  Vidal, C. \emph{Phys. Lett. A} {\bf77}, 391 (1980).
\bibitem{poincare4}Roux, J.C. \& Swinney, H. in Nonlinear Phenomena in Chemical Dynamics, edited by C. Vidal and A. Pacault (Springer-Verlag, Berlin, 1981).
\bibitem{poincare5} Takens, F., Dynamical Systems and Turbulence, Warwick, 1980, edited by D. A. Rand and L.-S. Young, Lecture Notes in Math Vol. 898 (Springer-Verlag, Berlin, 1981).
\bibitem{poincare6} Castro R., \& Sauer, T. Correlation dimension of attractors through interspike intervals. \emph{Phys. Rev. E} {\bf 55}, 287--290 (1997).


\bibitem{friston1995}Friston, K.J., Frith, C.D., Turner, R. , Frackowiak, R.S.J. Characterizing evoked hemodynamics with fMRI. \emph{Neuroimage} {\bf 2},157--165 (1995).

\bibitem{friston1998}Friston, K.J.,  Fletcher, P., Josephs, O., Holmes, A., Rugg, M.D., Turnera, R. Event-related fMRI: characterizing differential responses. \emph{Neuroimage} {\bf 7}, 30--40 (1998).

\bibitem{aguirre} Aguirre, G.K., Zarahn, E., D'Esposito., M.The variability of human BOLD hemodynamic responses. \emph{NeuroImage} {\bf 8}, 360--369 (1998).





\bibitem{bak}Bak, P. (1996) How nature works: The science of self-organized criticality (Copernicus Books, New York).

\bibitem{jensen}Jensen, H. J. (1998) Self-organized criticality (Cambridge University Press).

\bibitem{chialvo2010}Chialvo, D.R. Emergent complex neural dynamics. \emph{Nature Physics} {\bf 6}, 744--750 (2010).
\bibitem{expert} Expert, P., Lambiotte, R., Chialvo, D.R., Christensen, K., Jensen, H.J., Sharp, D.J.,  Turkheimer, F. Self-similar correlation function in brain resting state fMRI.  \emph{J. R. Soc. Interface} \textbf{8}, 472--479 (2011).




\bibitem{tononi} Tononi, G., Sporns, O.,  Edelman, G.M., A measure for brain complexity: relating functional  segregation and integration in the nervous system. \emph{Proc Natl Acad Sci USA} {\bf 91}, 5033--5037 (1994).

\bibitem{sporns}Sporns, O. (2010) Networks of the Brain, (MIT Press).

\bibitem{plenz}Beggs, J.M. \& Plenz, D. Neuronal avalanches in neocortical circuits. \emph{J. Neuroscience} \textbf{23}, 11167--11177, (2003).  

\bibitem{peterman}Petermann, T., Thiagarajan, TC., Lebedev, M.A., Nicolelis, M.A.L., Chialvo, D.R., Plenz D. Spontaneous cortical activity in awake monkeys composed of neuronal avalanches
\emph{Proc. Natl. Acad. Sci. U. S. A}  {\bf 106}, 15921--15926 (2009).




\bibitem{jensen2007} Vedel Jensen, E.B. \&  Thorarinsdottir, T.L.  A spatio-temporal model for functional magnetic resonance imaging data with a view to resting state networks. \emph{Scand. J. of Stat.} {\bf 34}, 587--614 (2007).

\bibitem{plenz3}Thiagarajan, T.C., Lebedev, M.A., Nicolelis, M.A., Plenz, D. Coherence potentials: loss-less, all-or-none network events in the cortex. \emph{PLoS Biol} {\bf 8} e1000278 (2010).


\bibitem{fraiman2010} Fraiman, D. \&  Chialvo, D.R.  Optimal information-sharing in brain resting state networks.  \emph{arXiv:1011.1192}  (2010)

\bibitem{koenig} Koenig T., Prichep L., Lehmann D., Valdes Sosa P., Braeker E., Kleinlogel H., Isenhart R., John E.R.  Millisecond by Millisecond, Year by Year: Normative EEG Microstates and Developmental Stages.  \emph{Neuroimage} \textbf{16}, 41--48 (2002).

\bibitem{kaplan} Kaplan A.Ya., Fingelkurts An.A., Fingelkurts Al.A., Borisov S.V., Darkhovsky B.S. Nonstationary nature of the brain activity as revealed by EEG/MEG: Methodological, practical and conceptual challenges. \emph{Signal Processing} \textbf{85}, 2190--2212 (2005).

\bibitem{vandeville} Van De Ville D., Britz J., Michel C.M. EEG microstate sequences in healthy humans at rest reveal scale-free dynamics. \emph{Proc. Natl. Acad. Sci. U. S. A}  {\bf 107}, 18179--18184 (2010).

\bibitem{allegrini} Allegrini P., Paradisi P., Menicucci D.,  Gemignani A. Fractal complexity in spontaneous EEG metastable-state transitions: new vistas on integrated neural dynamics. \emph{Frontiers in Physiology} {\bf 1}, doi:10.3389/fphys.2010.00128 (2010).


\end{thebibliography}

\begin{acknowledgments}
 Work supported by NIH (USA), CONICET (Argentina) and the MCyT (Spain). 
\end{acknowledgments}

\end{article}
\end{document}